%% file: ieee_paper.tex
\documentclass[10pt,journal,final]{IEEEtran}
\usepackage[T1]{fontenc}
\usepackage[utf8]{inputenc}
\usepackage{graphicx}
\usepackage{amsmath,amssymb,bm}
\usepackage[caption=false,font=footnotesize]{subfig}
\usepackage{booktabs,array}
\usepackage{stfloats}
\usepackage[noadjust]{cite}
\usepackage{microtype}
\usepackage{url}
\usepackage[hidelinks,breaklinks=true]{hyperref}
\makeatletter
\renewcommand{\section}{\@startsection{section}{1}{\z@}%
{6pt}{2pt}%
{\normalfont\normalsize\centering\scshape}}
\renewcommand{\subsection}{\@startsection{subsection}{2}{\z@}%
{4pt}{2pt}%
{\normalfont\normalsize\itshape}}
\makeatother
\newenvironment{compactequation}{%
\setlength{\abovedisplayskip}{4pt plus 1pt minus 1pt}
\setlength{\belowdisplayskip}{4pt plus 1pt minus 1pt}
\setlength{\abovedisplayshortskip}{0pt plus 1pt}
\setlength{\belowdisplayshortskip}{4pt plus 1pt minus 1pt}
\setlength{\jot}{1pt}
\begin{equation}}{\end{equation}}
\begin{document}

\title{DecompJSCC: Joint Source-Channel Coding With Decoupled SNR and Rate Adaptation}

\author{Yunhao Liu, \emph{Graduate Student Member, IEEE}, Jiancun Fan, \emph{Senior Member, IEEE},\\ Haojun Sun and Jie Luo%
\thanks{\textit{}}}
\IEEEaftertitletext{\vspace{-2\baselineskip}}
\maketitle

\begin{abstract}
Deep joint source-channel coding (DeepJSCC) for image transmission must adapt to changes in the signal-to-noise ratio (SNR) and available transmission resources. However, some existing joint-adaptive methods condition the encoder on the requested rate, while others apply static masks after encoding. The former couples latent formation to the rate request, whereas the latter cannot adapt channel selection to image content. To address these limitations, we propose DecompJSCC, which decomposes joint adaptation into SNR-conditioned latent formation and post-encoding rate control within a single model. The encoder applies SNR-conditioned block attention residuals across all stages, aggregating intra-stage and cross-stage semantic representations. At the receiver, an adapter performs SNR-conditioned affine correction. We introduce a post-encoding rate-adaptive channel semantic budgeting (RCSB) module whose shared gate selects channels of the formed latent by image content, adjusts their gains, and keeps exactly the requested count. This keeps the encoder output independent of the rate request for a fixed image and SNR. Experimental results show that DecompJSCC outperforms SwinJSCC and SA-RA-JSCC in peak signal-to-noise ratio (PSNR) across a range of SNR and rate settings, by 0.64 dB and 1.00 dB over SwinJSCC and by 0.97 dB and 0.96 dB over SA-RA-JSCC on Kodak and CLIC2021, while using only 7.5\% of SwinJSCC's Adaptive ModNet parameters.
\end{abstract}


\section{Introduction}
\label{sec:introduction}

Deep joint source-channel coding (DeepJSCC) learns an end-to-end mapping that sends images directly as channel symbols and reconstructs them at the receiver~\cite{bourtsoulatze2019deep}. In wireless links, the signal-to-noise ratio (SNR) and the available rate vary over time, and covering these conditions requires multiple models, which are costly to train and deploy. Therefore, joint adaptation with a single model is a central goal for image transmission.

Prior work achieves SNR adaptation through receiver-side conditional decoding~\cite{ding2021snr}, feature attention modulation~\cite{xu2022wireless}, and SNR embeddings and queries~\cite{zhang2025snreqjscc}. Rate adaptation, in turn, relies on progressive transmission~\cite{kurka2021bandwidth} and rate control driven by image content and channel conditions~\cite{yang2022deep}. For joint adaptation, SwinJSCC~\cite{yang2024swinjscc} employs SNR modulation modules and encoder-side rate modulation, STARJSCC~\cite{li2025star} combines channel-conditioned modulation with a post-encoding static mask, and SA-RA-JSCC~\cite{zhang2026sarjsc} performs global reweighting and channel selection on the encoded latent features. These joint designs show that the SNR and rate conditions can be injected at different points in the model.

However, encoder-side rate modulation~\cite{yang2024swinjscc} ties latent formation to the transmission budget, so changing the requested rate can alter both the encoded representation and the transmitted channels. Static post-encoding masking avoids this dependence but fixes the channel selection order across images, preventing the selected subset from adapting to the content of the current latent~\cite{li2025star},~\cite{zhang2026sarjsc}. A remaining design challenge is to combine SNR-conditioned latent formation with content-dependent channel selection while keeping the rate request outside the encoding process.

To address this challenge, we propose DecompJSCC, an image JSCC framework that decouples SNR-conditioned latent formation from post-encoding rate adaptation. SNR-conditioned block attention residuals aggregate representations across block depths and non-adjacent stages throughout the encoder, while an affine adapter provides SNR-conditioned correction at the receiver. The proposed rate-adaptive channel semantic budgeting (RCSB) module then uses a shared gate to score and modulate latent channels according to image content and the requested rate, with hard top-$k$ selection enforcing the channel budget exactly. The rate request is thus confined to post-encoding processing, leaving the encoder output unchanged for a fixed image, SNR, and model state.

\section{System Model}

\label{sec:system}

\def\FigOneSplitLayout{1}
\input{fig/fig01_layout.tex}
\let\FigOneSplitLayout\undefined

We consider point-to-point wireless image transmission under varying channel conditions and payload budgets. The nominal SNR $\gamma$ is expressed in dB and assumed known to both ends, while the requested channel count $k$ specifies the payload budget. The source image $\mathbf x\in\mathbb R^{H\times W\times C}$ has pixel values normalized to $[0,1]$, where $H$, $W$, and $C$ are its height, width, and number of color channels, with $C=3$ for RGB images.

Following the signal flow in Fig.~\ref{fig:framework}(a), the SNR-conditioned AttnRes encoder $F_\theta$, with trainable parameters $\theta$, forms the latent $\mathbf z_\gamma=F_\theta(\mathbf x,\gamma)\in\mathbb R^{N\times D}$, where $N$ and $D$ are the numbers of latent tokens and feature channels. To accommodate the requested payload budget, the post-encoding RCSB module $\mathcal R_\eta$, with trainable parameters $\eta$, selects and reweights the feature channels of $\mathbf z_\gamma$ for a requested count $k\in\mathcal K\subseteq\{1,\ldots,D\}$. Specifically, $(\mathbf z_{\mathrm{tx},k},\mathbf m_k)=\mathcal R_\eta(\mathbf z_\gamma,k)$, where $\mathbf z_{\mathrm{tx},k}$ denotes the transmit latent and $\mathbf m_k$ the binary feature-channel selection mask. The encoder $F_\theta$ does not take $k$ as an input. The mask $\mathbf m_k\in\{0,1\}^{1\times D}$ selects the same $k$ feature channels for all tokens, so $\|\mathbf m_k\|_0=k$, where $\|\cdot\|_0$ counts nonzero entries. The transmit latent $\mathbf z_{\mathrm{tx},k}\in\mathbb R^{N\times D}$ retains the dimensions of $\mathbf z_\gamma$, with unselected channels set to zero.

The shared mask selects $k$ real coefficients from each of the $N$ tokens, giving $Nk$ coefficients in total. Their mean squared amplitude is $P_{\mathrm{act},k}=\|\mathbf z_{\mathrm{tx},k}\|_{\mathrm F}^2/(Nk)$, where $\|\cdot\|_{\mathrm F}$ denotes the Frobenius norm. Let $\mathcal Q:\mathbb R^{N\times D}\to\mathbb C^{ND/2}$ map each pair of real coefficients to a complex symbol. Power normalization and pairing yield the complex channel input vector $\mathbf p_k=\mathcal Q(\mathbf z_{\mathrm{tx},k}/\sqrt{2P_{\mathrm{act},k}})$, with total energy $\|\mathbf p_k\|_2^2=Nk/2$. The nominal payload channel bandwidth ratio (CBR) is the ratio of $Nk/2$ complex payload symbols to $HWC$ source components, giving $\mathrm{CBR}_{\mathrm{payload}}=Nk/(2HWC)$.

Over the wireless channel, the received signal is $\mathbf y_k^{\mathrm{ch}}=\mathbf h\odot\mathbf p_k+\mathbf n_\gamma$, where $\mathbf h$ is the channel-gain vector and $\odot$ denotes elementwise multiplication. We consider an additive white Gaussian noise (AWGN) channel with $\mathbf h=\mathbf1$ and $\mathbf n_\gamma\sim\mathcal{CN}(\mathbf0,10^{-\gamma/10}\mathbf I)$.

With $\mathbf m_k$ and $P_{\mathrm{act},k}$ assumed known, the receiver converts the received symbols back to real coefficients, restores their scale, and reapplies the mask to obtain the received latent $\widetilde{\mathbf y}_k=\sqrt{2P_{\mathrm{act},k}}\,\mathcal Q^{-1}(\mathbf y_k^{\mathrm{ch}})\odot\mathbf m_k\in\mathbb R^{N\times D}$. The SNR-conditioned adapter $C_\omega$ then applies an affine correction to $\widetilde{\mathbf y}_k$, and the JSCC decoder $G_\phi$ produces the raw reconstruction $\widehat{\mathbf x}_{\gamma,k}=G_\phi(C_\omega(\widetilde{\mathbf y}_k,\gamma))\in\mathbb R^{H\times W\times C}$. Here, $\omega$ and $\phi$ are the trainable parameters of the adapter and decoder, respectively.

Because the raw reconstruction may contain pixel values outside $[0,1]$, it is clipped elementwise to this range before computing the mean squared error (MSE) during training and evaluation. The model is trained to minimize $\mathcal L=\mathbb E[\|\mathbf x-\operatorname{clip}_{[0,1]}(\widehat{\mathbf x}_{\gamma,k})\|_{\mathrm F}^2/(HWC)]$, where the expectation is taken over source images, sampled SNRs and channel counts, and channel noise.

\section{Proposed DecompJSCC}

\label{sec:method}

Building on the system model, DecompJSCC uses SNR-conditioned depth aggregation to form the latent representation and post-encoding RCSB to select and reweight its feature channels. An SNR-conditioned affine adapter then corrects the received latent before decoding.

\subsection{SNR-Conditioned Block Attention Residuals}

Within a hierarchical Swin encoder~\cite{yang2024swinjscc},~\cite{liu2021swin}, we extend Block AttnRes~\cite{kimi2026attnres} with SNR conditioning to aggregate representations across depths at each token position. In Fig.~\ref{fig:framework}(b), stage $t$ contains $L_t$ Swin--AttnRes blocks with $N_t$ tokens and $D_t$ feature channels. Aggregation precedes the attention and MLP branches of each block, with an additional aggregation after the last block, giving $2L_t+1$ mixing points. Let $\mathbf h_{t,\ell}\in\mathbb R^{N_t\times D_t}$ denote the output at mixing point $\ell$. Given stage input $\mathbf b_{t,0}$, block $n$ computes the attention output $\mathbf a_{t,n}=f^{\mathrm{attn}}_{t,n}(\mathbf h_{t,2n-1})$ and block output $\mathbf b_{t,n}=\mathbf a_{t,n}+f^{\mathrm{mlp}}_{t,n}(\mathbf h_{t,2n})$. Here $f^{\mathrm{attn}}_{t,n}$ and $f^{\mathrm{mlp}}_{t,n}$ apply window attention and an MLP, respectively, each with its preceding normalization. At mixing point $\ell=2n-1$ before attention, the local sources are $\{\mathbf b_{t,0},\ldots,\mathbf b_{t,n-1}\}$. Before the MLP, mixing point $\ell=2n$ additionally includes $\mathbf a_{t,n}$. At the stage-output mixing point $\ell=2L_t+1$, the local sources are $\{\mathbf b_{t,0},\ldots,\mathbf b_{t,L_t}\}$. Between stages, the aggregated output passes through PatchMerging. The final stage's aggregated output is normalized to form $\mathbf z_\gamma$.

In addition to local sources, stages $t\ge3$ incorporate the aggregated outputs $\mathbf h_{s,2L_s+1}$ of earlier stages $s=1,\ldots,t-2$. To align these outputs with the spatial resolution and feature width of stage $t$, the alignment mapping $\mathcal T_{s\to t}$ reuses the corresponding PatchMerging chain. The resulting cross-stage sources $\mathcal T_{s\to t}(\mathbf h_{s,2L_s+1})$ are shared across all mixing points within stage $t$. Stages 1 and 2, however, use only local sources.

For either the local or cross-stage branch, let $\mathcal S$ denote the index set of $M$ source tensors $\mathbf v_s\in\mathbb R^{N_t\times D_t}$, ordered from shallow to deep. To condition source weighting on deviations from a reference SNR $\gamma_{\mathrm{ref}}$, we use $\Delta\gamma=(\gamma-\gamma_{\mathrm{ref}})/\tau_\gamma$ with a fixed scaling factor $\tau_\gamma>0$. For each source $s\in\mathcal S$, the tokenwise scores $\mathbf e_s$ and normalized weights $\boldsymbol\alpha_s$ are computed as
\begin{compactequation}
\begin{aligned}
\mathbf e_s
&=\frac{\operatorname{RMSNorm}(\mathbf v_s)\mathbf q_{t,\ell}}{\sqrt{D_t}}
+\beta_{t,\ell}r_s\Delta\gamma\,\mathbf1_{N_t},\\
\boldsymbol\alpha_s
&=\frac{\exp(\mathbf e_s)}{\sum_{j\in\mathcal S}\exp(\mathbf e_j)}.
\end{aligned}
\label{eq:score}
\end{compactequation}
At each mixing point, the two branches share the learnable query $\mathbf q_{t,\ell}\in\mathbb R^{D_t\times1}$ and SNR-bias coefficient $\beta_{t,\ell}\in\mathbb R$, while applying softmax separately over their sources at each token. The ranks $r_s$ introduce depth dependence into the SNR bias and are evenly spaced from $-1$ to $1$ in source order within each set, with $r_s=0$ for a single source. RMSNorm acts along feature channels for scoring only, and $\mathbf1_{N_t}$ denotes an all-ones vector.

Let $\mathcal S^{\mathrm{loc}}_{t,\ell}$ index the local source tensors $\mathbf v_s^{\mathrm{loc}}$, and let $\mathbf v_s^{\mathrm{cross}}=\mathcal T_{s\to t}(\mathbf h_{s,2L_s+1})$ denote an aligned cross-stage source. Using source values before scoring normalization, the two branches form the local aggregate $\mathbf h^{\mathrm{loc}}_{t,\ell}=\sum_{s\in\mathcal S^{\mathrm{loc}}_{t,\ell}}\boldsymbol\alpha_s^{\mathrm{loc}}\odot\mathbf v_s^{\mathrm{loc}}$ and the cross-stage aggregate $\mathbf h^{\mathrm{cross}}_{t,\ell}=\sum_{s=1}^{t-2}\boldsymbol\alpha_s^{\mathrm{cross}}\odot\mathbf v_s^{\mathrm{cross}}$, with weights broadcast across feature channels. For $t\le2$, $\mathbf h^{\mathrm{cross}}_{t,\ell}=\mathbf0$. Combining these aggregates gives the mixing-point output
\begin{compactequation}
\mathbf h_{t,\ell}
=\mathbf h^{\mathrm{loc}}_{t,\ell}
+\tanh(\lambda_{t,\ell})\mathbf h^{\mathrm{cross}}_{t,\ell}.
\label{eq:mixture}
\end{compactequation}
The learnable scalar $\lambda_{t,\ell}$ controls cross-stage injection. We initialize $\mathbf q_{t,\ell}$, $\beta_{t,\ell}$, and $\lambda_{t,\ell}$ to zero, yielding uniform initial source weights and zero cross-stage contribution.

\subsection{Post-Encoding RCSB}

After SNR-conditioned encoding, RCSB $\mathcal R_\eta$ uses a shared gate to select $k$ feature channels and modulate their gains, as shown in Fig.~\ref{fig:framework}(c). $\operatorname{FeatureProj}$ maps each token of $\mathbf z_\gamma$ from $D$ to $D_{\mathrm h}$ dimensions, while $\operatorname{RateMod}$ maps $k$ to a $1\times D_{\mathrm h}$ modulation vector. Both mappings use MLPs, with the rate vector broadcast across tokens. The elementwise product of their outputs forms the rate-conditioned control features $\mathbf f_k\in\mathbb R^{N\times D_{\mathrm h}}$, which a linear $\operatorname{GateHead}$ and sigmoid $\sigma$ map to the shared gate $\mathbf g_k\in(0,1)^{N\times D}$ as follows
\begin{compactequation}
\begin{aligned}
\mathbf f_k
&=\operatorname{FeatureProj}\!\left(\operatorname{stopgrad}(\mathbf z_\gamma)\right)
\odot\operatorname{RateMod}(k),\\
\mathbf g_k
&=\sigma\!\left(\operatorname{GateHead}(\mathbf f_k)\right).
\end{aligned}
\label{eq:rcsb}
\end{compactequation}
We first average $\mathbf g_k$ over tokens to obtain the channel scores $\mathbf s_k=N^{-1}\mathbf1_N^{\mathsf T}\mathbf g_k\in\mathbb R^{1\times D}$, then select the $k$ highest-scoring channels using the mask $\mathbf m_k=\operatorname{HardTopK}(\mathbf s_k,k)$. With this mask, RCSB normalizes $\mathbf g_k$ using its root mean square $r_{\mathrm{act},k}$ over the $Nk$ selected entries to obtain the gains $\widetilde{\mathbf g}_k$ as follows
\begin{compactequation}
r_{\mathrm{act},k}=\sqrt{\frac{\|\mathbf g_k\odot\mathbf m_k\|_{\mathrm F}^2}{Nk}},\quad
\widetilde{\mathbf g}_k=\frac{\mathbf g_k}{\max(r_{\mathrm{act},k},\epsilon)},
\label{eq:active_rms}
\end{compactequation}
where $\epsilon=10^{-6}$ ensures numerical stability. RCSB forms the transmit latent $\mathbf z_{\mathrm{tx},k}=\mathbf z_\gamma\odot\mathbf m_k\odot\widetilde{\mathbf g}_k$ before the power normalization described in Section~\ref{sec:system}.

To stabilize encoder training, $\operatorname{stopgrad}$ blocks gradients from the control branch to the encoder without changing forward values. Reconstruction gradients still update the encoder through $\mathbf z_\gamma$, while the shared gate learns through gain modulation with the $\operatorname{HardTopK}$ mask held fixed during backpropagation.

\subsection{Receiver-Side SNR Correction}

At the receiver, $C_\omega$ adjusts $\widetilde{\mathbf y}_k$ before decoding, as shown in Fig.~\ref{fig:framework}(a). A two-layer MLP takes the scaled SNR $\gamma/\tau_\gamma$ as input and produces two vectors $\boldsymbol\rho_\omega(\gamma),\boldsymbol\delta_\omega(\gamma)\in\mathbb R^{1\times D}$ to control feature scaling and shifting, respectively. These vectors are shared across tokens, giving the affine correction
\begin{compactequation}
\begin{aligned}
C_\omega(\widetilde{\mathbf y}_k,\gamma)
&=\widetilde{\mathbf y}_k\odot[\mathbf1+0.1\tanh(\boldsymbol\rho_\omega(\gamma))]\\
&\quad+0.1\tanh(\boldsymbol\delta_\omega(\gamma)).
\end{aligned}
\label{eq:adapter}
\end{compactequation}
With the MLP output layer initialized to zero, the adapter initially acts as an identity map. Finally, $G_\phi$ decodes the corrected features to reconstruct the image.

\setcounter{topnumber}{2}
\section{Experimental Results}
\label{sec:experiments}

\begin{figure}[!t]
\centering
\subfloat[]{\includegraphics[width=0.49\columnwidth,trim=0bp 2.5bp 0bp 0.75bp,clip]{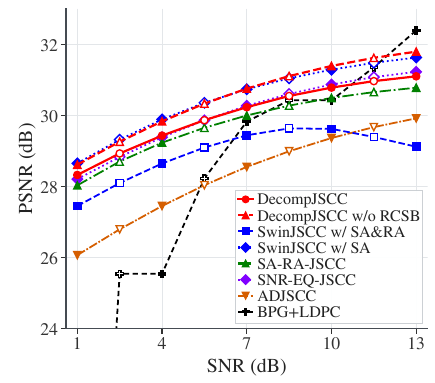}}\hfill
\subfloat[]{\includegraphics[width=0.49\columnwidth,trim=0bp 2.5bp 0bp 0.75bp,clip]{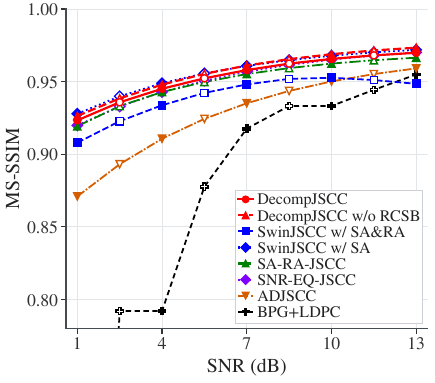}}
\par\vspace{1pt}
\subfloat[]{\includegraphics[width=0.49\columnwidth,trim=0bp 2.5bp 0bp 0.75bp,clip]{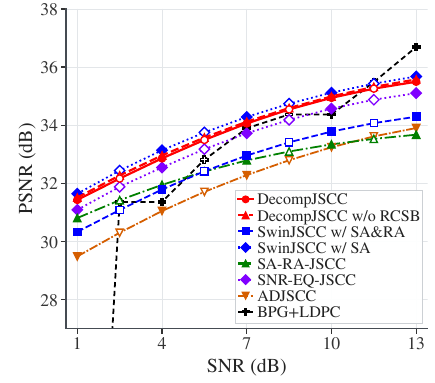}}\hfill
\subfloat[]{\includegraphics[width=0.49\columnwidth,trim=0bp 2.5bp 0bp 0.75bp,clip]{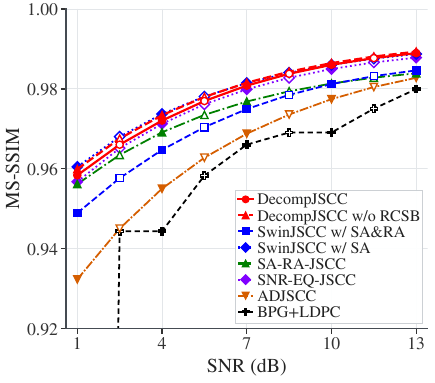}}
\par\vspace{1pt}
\subfloat[]{\includegraphics[width=0.49\columnwidth,trim=0bp 2.5bp 0bp 0.75bp,clip]{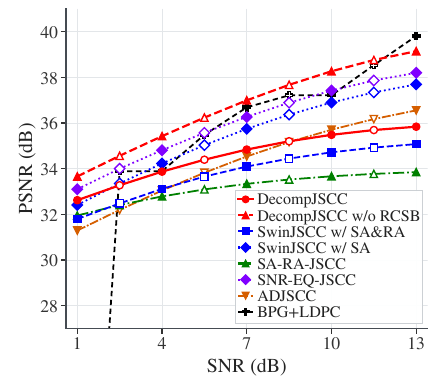}}\hfill
\subfloat[]{\includegraphics[width=0.49\columnwidth,trim=0bp 2.5bp 0bp 0.75bp,clip]{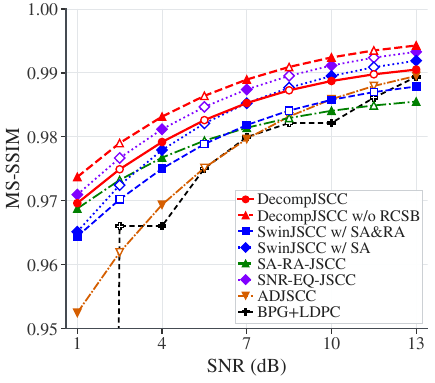}}
\caption{PSNR and MS-SSIM versus SNR on CLIC2021 over AWGN channels, with CBRs of $1/48$, $1/16$, and $1/8$ from top to bottom.}
\label{fig:snr}
\end{figure}

\begin{figure}[!t]
\centering
\subfloat[]{\includegraphics[width=0.49\columnwidth,trim=0bp 3.75bp 0bp 2.75bp,clip]{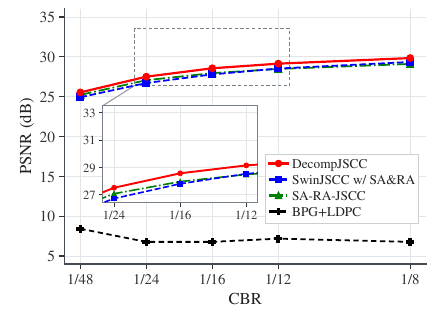}}\hfill
\subfloat[]{\includegraphics[width=0.49\columnwidth,trim=0bp 3.75bp 0bp 2.75bp,clip]{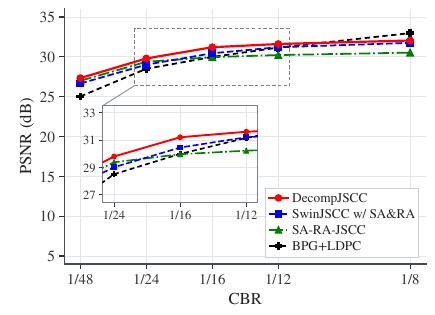}}
\par\vspace{1pt}
\subfloat[]{\includegraphics[width=0.49\columnwidth,trim=0bp 3.75bp 0bp 2.75bp,clip]{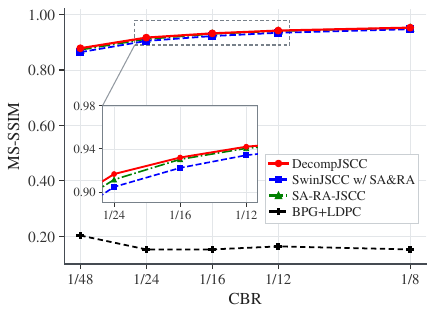}}\hfill
\subfloat[]{\includegraphics[width=0.49\columnwidth,trim=0bp 3.75bp 0bp 2.75bp,clip]{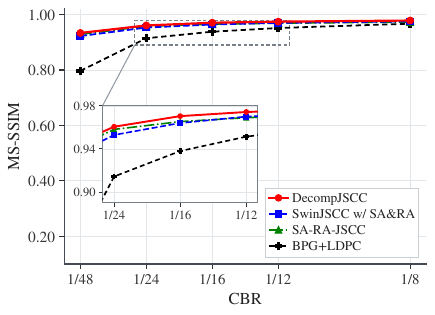}}
\caption{PSNR and MS-SSIM versus CBR on Kodak over AWGN channels at SNRs of 1 dB in (a),(c) and 7 dB in (b),(d).}
\label{fig:rate}
\end{figure}

\begin{figure}[!tbp]
\centering
\subfloat[]{\includegraphics[width=\columnwidth,trim=0bp 1bp 0bp 0.5bp,clip]{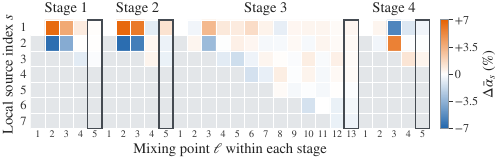}}
\par\vspace{1pt}
\subfloat[]{\includegraphics[width=\columnwidth,trim=0bp 1bp 0bp 0.5bp,clip]{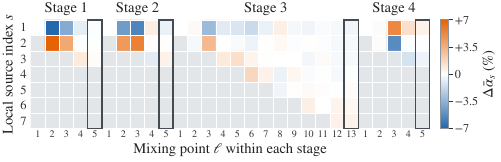}}
\caption{SNR-dependent local aggregation on Kodak at CBR=$1/16$. Heatmaps show changes in mean source weights from 7 dB to (a) 1 dB and (b) 13 dB, expressed in percentage points. Gray denotes unavailable sources.}
\label{fig:aggregation}
\end{figure}

\begin{figure*}[!tbp]
\centering
\includegraphics[width=0.92\textwidth,trim=0bp 1.75bp 0bp 2bp,clip]{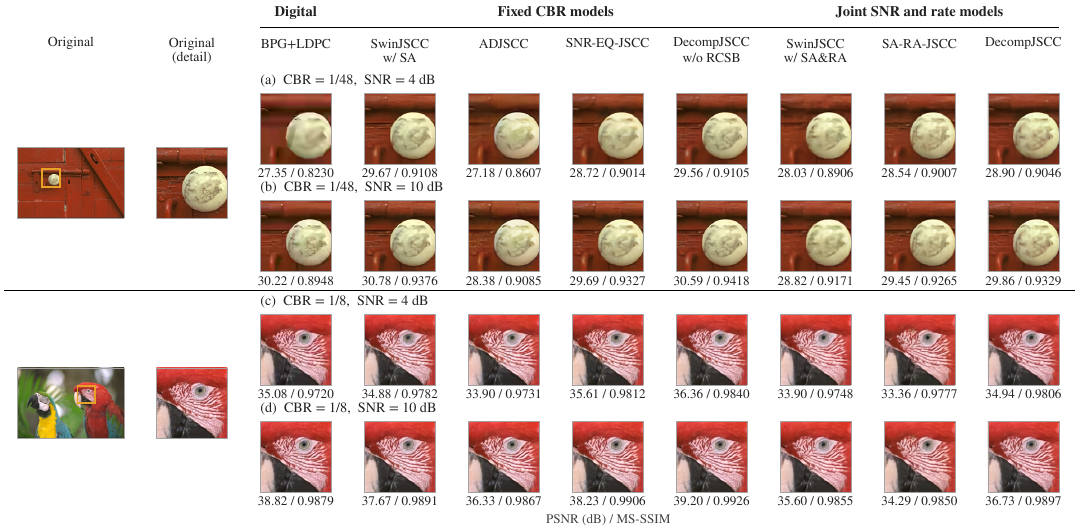}
\caption{Visual comparison of reconstructed Kodak images. Yellow boxes indicate the $128\times128$ regions enlarged for comparison. The values below each crop give PSNR (dB) and MS-SSIM for the corresponding full reconstruction.}
\label{fig:visual}
\end{figure*}

\subsection{Experimental Setup}

In our experiments, the proposed DecompJSCC model was trained and evaluated over AWGN channels. Training used $256\times256$ patches randomly cropped from the CLIC and DIV2K datasets. For evaluation, Kodak and CLIC2021 images were centrally cropped to dimensions divisible by 128 without resizing.

We jointly optimized the pretrained SwinJSCC w/o SA\&RA backbone~\cite{yang2024swinjscc} and our adaptation modules using Adam to minimize MSE, with pretraining conducted at a fixed SNR and CBR. Training covered $\gamma\in\{1,4,7,10,13\}$ dB and $k\in\{32,64,96,128,192\}$. The final $H/16\times W/16$ latent gives a payload CBR of $k/1536$ for RGB images. Setting the depth bias to $\gamma_{\mathrm{ref}}=7$ dB, the training midpoint, we also evaluated interpolation at the four midpoints between adjacent training SNRs.

The neural baselines include ADJSCC~\cite{xu2022wireless} and SNR-EQ-JSCC~\cite{zhang2025snreqjscc}, each trained separately for every CBR, and SwinJSCC w/ SA\&RA~\cite{yang2024swinjscc} and SA-RA-JSCC~\cite{zhang2026sarjsc}, each covering all evaluated CBRs with one model. We also compared DecompJSCC w/o RCSB against SwinJSCC w/ SA for SNR adaptation at fixed CBRs, training a separate model per method and CBR. Our implementations of ADJSCC, SNR-EQ-JSCC, and SA-RA-JSCC followed their published descriptions, with all neural models using MSE as the reconstruction loss.

Alongside these neural baselines, we included a digital baseline combining BPG compression with 5G NR LDPC coding and quadrature amplitude modulation (QAM). It uses information blocks of 8352 bits, with modulation and coding modes calibrated for each SNR on independent random bits and fixed before evaluation.

During testing, RGB reconstruction quality was measured by PSNR and multiscale structural similarity (MS-SSIM), with each metric averaged over three noise realizations per image and then equally across images. Failed BPG+LDPC transmissions were scored using black reconstructions.

\subsection{Reconstruction Performance}

We first compare reconstruction performance across SNRs and CBRs in Figs.~\ref{fig:snr} and~\ref{fig:rate}, respectively. DecompJSCC achieves higher PSNR and MS-SSIM than both joint neural baselines, including at SNRs not used during training. Across all the training SNRs and CBRs, DecompJSCC improves mean PSNR over SwinJSCC w/ SA\&RA by 0.64 dB on Kodak and 1.00 dB on CLIC2021. The corresponding gains over SA-RA-JSCC are 0.97 and 0.96 dB, respectively.

With one model covering all five CBRs, DecompJSCC nearly matches SNR-EQ-JSCC at CBR=$1/48$ and exceeds it by 0.35 dB at $1/16$, but trails it by 1.44 dB at $1/8$. When trained separately for each CBR, DecompJSCC w/o RCSB outperforms SwinJSCC w/ SA by 1.30 dB at $1/8$ but trails it by 0.15 dB at $1/16$.

DecompJSCC achieves consistently higher MS-SSIM than BPG+LDPC across Figs.~\ref{fig:snr} and~\ref{fig:rate} and higher PSNR at low SNRs, particularly at 1 dB, where decoding and budget failures sharply degrade the digital baseline's performance. At 13 dB, however, BPG+LDPC achieves the highest PSNR at all three CBRs in Fig.~\ref{fig:snr}.

\subsection{SNR-Conditioned Aggregation and RCSB}

The learned aggregation adjusts source weighting with SNR in both local and cross-stage branches. On Kodak at CBR=$1/16$, Fig.~\ref{fig:aggregation} plots the mean local weight differences relative to 7 dB, $\Delta\bar{\alpha}_s(\gamma)=\bar{\alpha}_s(\gamma)-\bar{\alpha}_s(7)$, revealing different responses across stages and mixing points. Mean cross-stage weights in Stage 4 change by up to 12.3 percentage points between 1 and 13 dB. We further evaluated the trained model with and without the SNR depth bias at test time, retaining the receiver adapter and using the same noise realization for each image in both settings. Without the SNR depth bias, mean PSNR decreased by 0.031 to 0.411 dB under both low and high SNR conditions.

At 7 dB and $k=32$, RCSB selected 32 feature channels per image, with 11 distinct channel sets across 18 Kodak images of the same dimensions, demonstrating selection based on image content. Across all budgets, the selected channel count matched the requested $k$, while each image's encoder latent remained unchanged.

\subsection{Visual Comparison}

To complement the quantitative results, we compare reconstructions of two Kodak images in Fig.~\ref{fig:visual}. DecompJSCC could better preserve local details than the two joint baselines, particularly around the parrot's eye, and achieves higher PSNR and MS-SSIM under all four conditions described below.

\subsection{Complexity Analysis}

We further compare model complexity in Table~\ref{tab:complexity} using $256\times256$ RGB inputs at CBR=$1/8$ and SNR=7 dB. DecompJSCC uses 26.5\% fewer parameters than SwinJSCC w/ SA\&RA with 1.4\% more counted FLOPs. DecompJSCC supports all evaluated CBRs with one model, whereas ADJSCC and SNR-EQ-JSCC each use five separately trained models. Its total FP32 parameter storage is consequently 54.3\% and 82.8\% lower, respectively.

\begin{table}[!tbp]
\caption{Model Complexity and Storage Requirements}
\label{tab:complexity}
\centering
\footnotesize
\setlength{\tabcolsep}{2pt}
\renewcommand{\arraystretch}{1.10}
\begin{tabular*}{\columnwidth}{@{\extracolsep{\fill}}lrrcr@{}}
\toprule
Method & Params. (M) & GFLOPs & Models & Storage (MiB) \\
\midrule
DecompJSCC & 24.320 & 72.426 & 1 & 92.775 \\
SwinJSCC w/ SA\&RA & 33.073 & 71.418 & 1 & 126.165 \\
SA-RA-JSCC & 24.032 & 69.014 & 1 & 91.674 \\
SNR-EQ-JSCC & 28.363 & 68.874 & 5 & 539.894 \\
ADJSCC & 10.707 & 93.384 & 5 & 202.848 \\
\bottomrule
\end{tabular*}
\par\vspace{2pt}
\begin{minipage}{\columnwidth}
\normalfont\footnotesize\raggedright
Parameters and FLOPs refer to one model, while model counts and FP32 parameter storage cover all CBRs. FLOPs include supported convolution and matrix operations, counting multiplication and addition separately.
\end{minipage}
\end{table}

Excluding the backbone, DecompJSCC uses 1.112 M active parameters for AttnRes aggregation, RCSB, and the receiver SNR adapter. This represents only 7.5\% of the combined 14.797 M parameters in SwinJSCC's ModNet modules for SNR and rate adaptation. DecompJSCC achieves better reconstruction quality than SwinJSCC w/ SA\&RA while using substantially fewer adaptation parameters.

\section{Conclusions}
\label{sec:conclusions}

We proposed DecompJSCC to decouple SNR-conditioned latent formation from post-encoding rate adaptation. Depth aggregation and receiver correction provide SNR adaptation, while RCSB selects and reweights latent channels according to image content and the requested budget. Experiments on Kodak and CLIC2021 over AWGN channels demonstrate gains in both PSNR and MS-SSIM across varying SNRs and CBRs. DecompJSCC achieves these gains across all evaluated CBRs using a single model and substantially fewer adaptation parameters than SwinJSCC w/ SA\&RA. Future work will explore extending DecompJSCC to other modalities and multimodal token communication.

\bibliographystyle{IEEEtran}
\bibliography{ieee_references}
\end{document}

%% file: fig/fig01_layout.tex
\ifdefined\FigOneSplitLayout
\begin{figure*}[!t]
\centering
\includegraphics[width=0.95\textwidth,trim=0bp 4.5bp 0bp 4.25bp,clip]{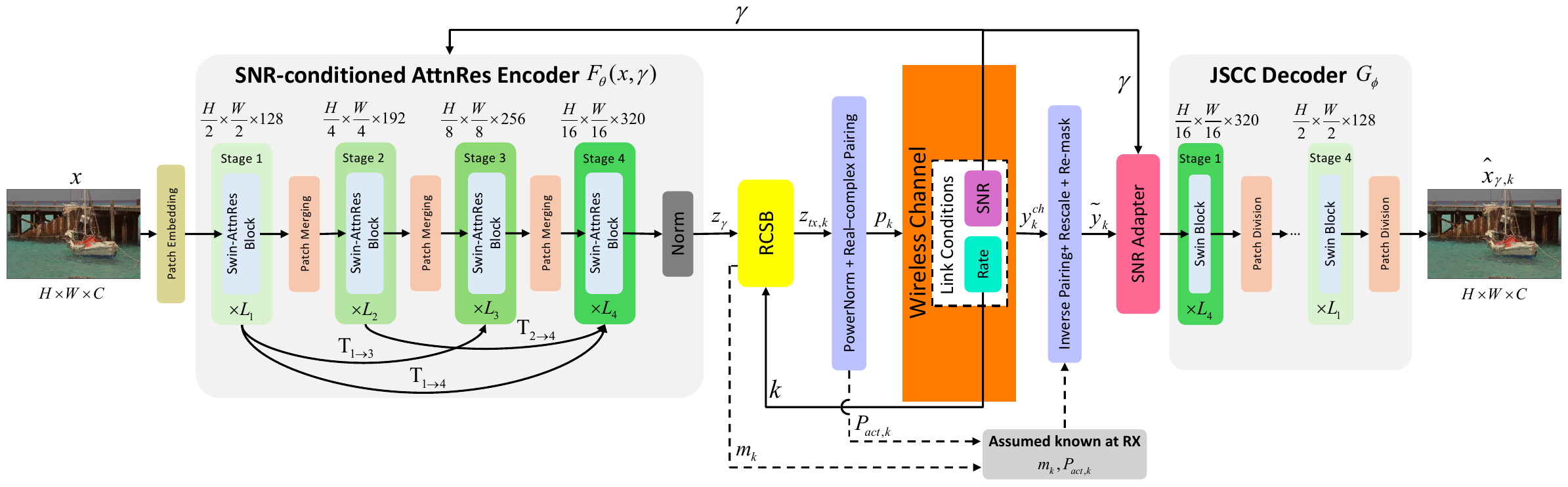}
\par{\footnotesize (a)}
\end{figure*}

\begin{figure}[t]
\centering
\includegraphics[width=1.00\columnwidth,trim=0bp 4.5bp 0bp 5.25bp,clip]{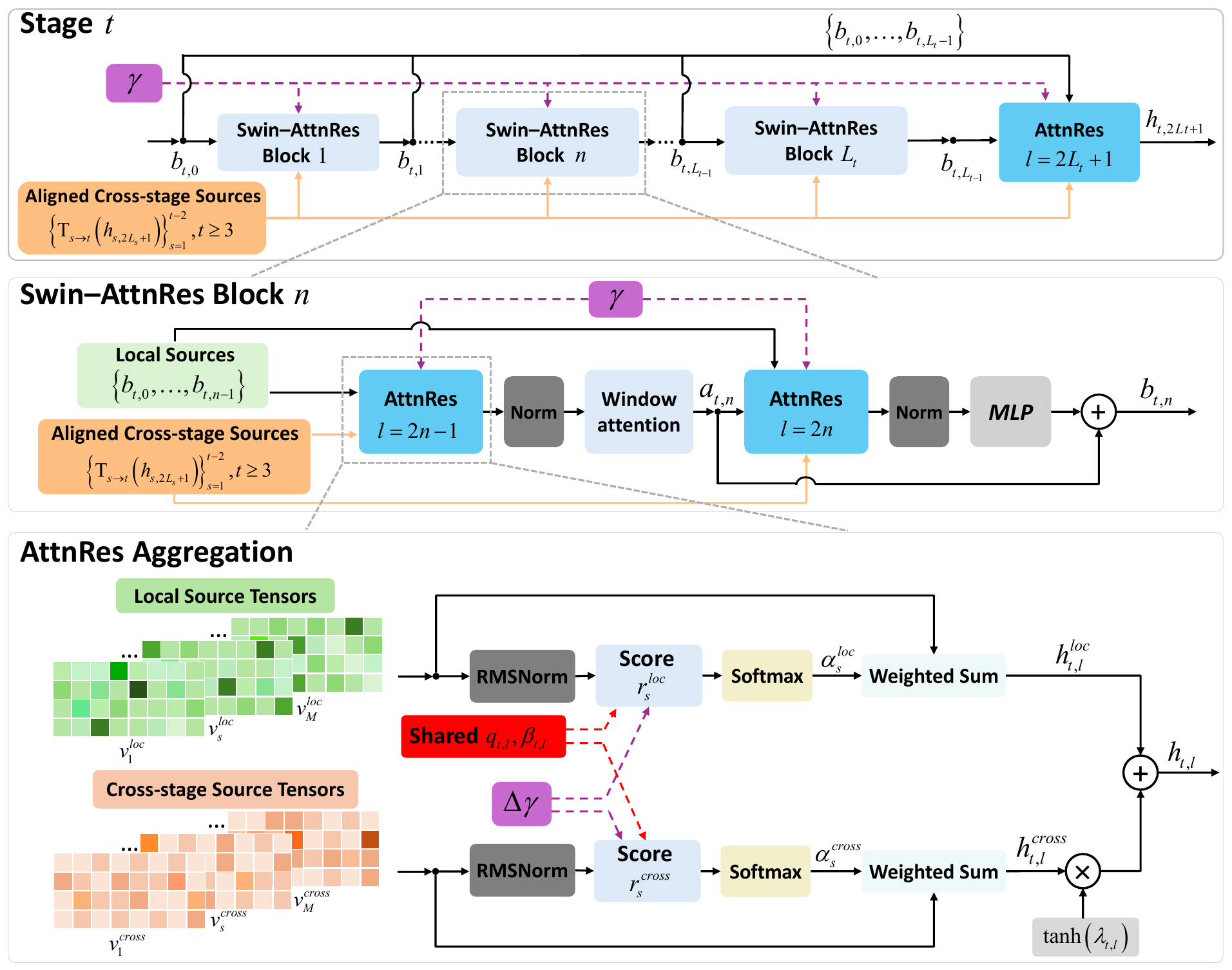}
\par{\footnotesize (b)}
\caption{Architecture of DecompJSCC.  (a) System model. (b) Swin--AttnRes stage and block structures with SNR-conditioned aggregation. (c) Post-encoding RCSB.}
\label{fig:framework}
\end{figure}

\begin{figure}[t]
\centering
\includegraphics[width=\columnwidth,trim=0bp 3.25bp 0bp 2.25bp,clip]{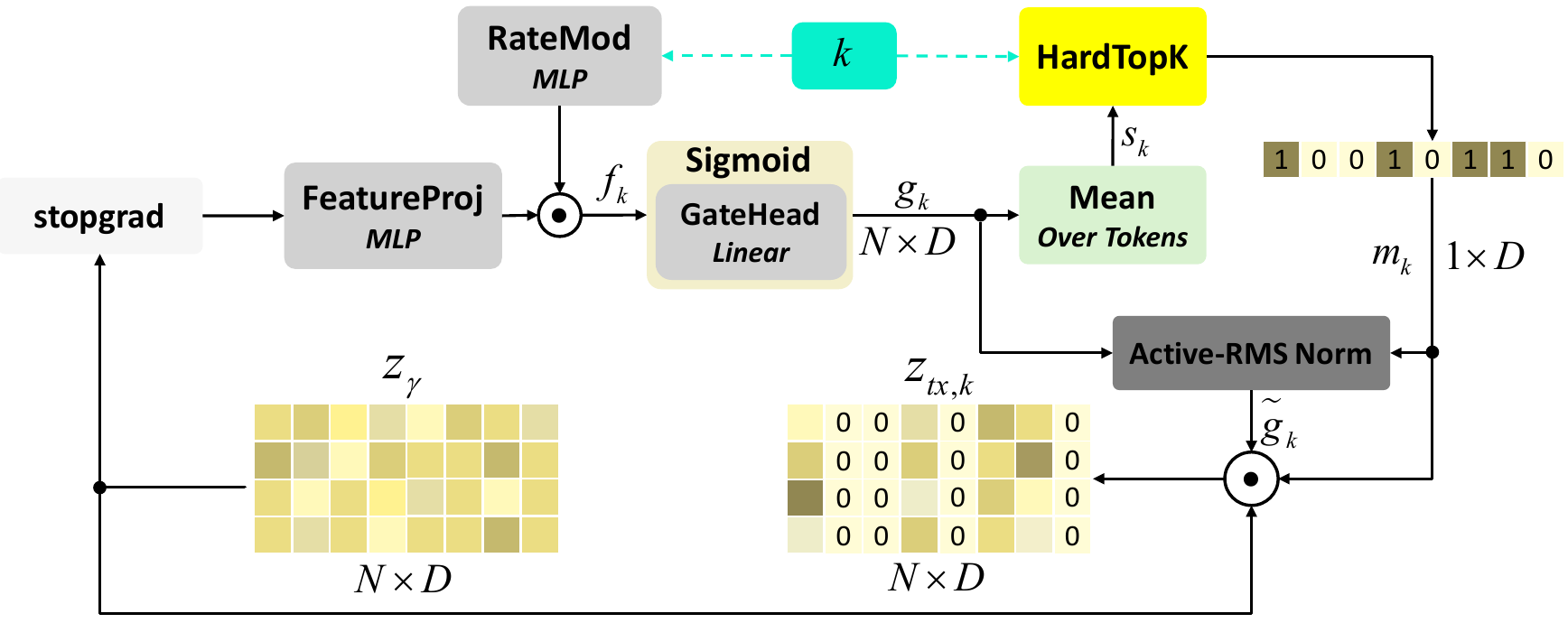}
\par{\footnotesize (c)}
\end{figure}
\else
\includegraphics[width=0.95\textwidth,trim=0bp 4.5bp 0bp 4.25bp,clip]{fig/fig01_a.pdf}
\par{\footnotesize (a)}\par\vspace{1pt}
\begin{minipage}[c]{0.46\textwidth}
\centering
\includegraphics[width=\linewidth,trim=0bp 4.5bp 0bp 5.25bp,clip]{fig/fig01_b.pdf}
\par{\footnotesize (b)}\par\vspace{1pt}
\end{minipage}\hfill
\begin{minipage}[c]{0.52\textwidth}
\centering
\includegraphics[width=\linewidth,trim=0bp 3.25bp 0bp 2.25bp,clip]{fig/fig01_c.pdf}
\par{\footnotesize (c)}\par\vspace{1pt}
\end{minipage}

\fi